\documentclass[preprint2]{aastex}

\begin{document}


\title{Abundant crystalline silicates in the disk of a very low mass star\footnote{Based partly on observations obtained at the ESO NTT at La Silla (Chile) for program 276.C\-5031. }}


\author{B. Mer\'{\i}n\altaffilmark{1}, 
	  J.-C. Augereau\altaffilmark{2},
	  E.  F. van Dishoeck\altaffilmark{1},
	  J. Kessler-Silacci\altaffilmark{3},
	  C. P. Dullemond\altaffilmark{4},
	  G. A. Blake,\altaffilmark{5},
	  F. Lahuis\altaffilmark{6,1},
	  J. M. Brown\altaffilmark{5},
	  V. C. Geers\altaffilmark{1},
	  K. M. Pontoppidan\altaffilmark{5},
	  F. Comer\'on\altaffilmark{7},
	  A. Frasca\altaffilmark{8},
	  S. Guieu\altaffilmark{2},
          J. M. Alcal\'a\altaffilmark{9},
	  A. C. A. Boogert\altaffilmark{10},
	  N. J. Evans, II\altaffilmark{3},
	  P. D'Alessio\altaffilmark{11},
	  L. G. Mundy\altaffilmark{12},
	  N. Chapman\altaffilmark{12}
} 

\altaffiltext{1}{Leiden Observatory, Leiden University, PO Box 9513, 
NL-2300 RA, Leiden.  The Netherlands}

\altaffiltext{2}{Laboratoire d'Astrophysique de Grenoble, 
	Universit\'e Joseph Fourier,  B.P. 53, 38041 Grenoble Cedex 9, 
	France }
 
\altaffiltext{3}{Department of Astronomy, University of Texas at Austin, 
	    1 University Station C1400 Austin, TX 78712-0259, USA}
 
\altaffiltext{4}{Max-Planck-Institut f\"ur Astronomie, K\"onigstuhl 17
	    D-69117, Heidelberg, Germany}

\altaffiltext{5}{Division of Geological and Planetary Sciences, MS 150-21, 
	    California Institute of Technology, Pasadena, CA 91125, USA}

\altaffiltext{6}{SRON Netherlands Institute for Space Research, PO Box 800, 
	    9700 AV Groningen, The Netherlands}

\altaffiltext{7}{European Southern Observatory, Karl-Schwarzschild-Strasse 2, 
85748 Garching bei M\"unchen, Germany}

\altaffiltext{8}{INAF - Osservatorio Astrofisico di Catania, via S. Sofia, 78,
 95123 Catania, Italy }

\altaffiltext{9}{INAF - Osservatorio Astronomico di Capodimonte, 
via Moiariello 16, I-80131, Naples, Italy}

\altaffiltext{10}{Division of Physics, Mathematics and Astronomy, MS 105-24,
	    California Institute of Technology, Pasadena, CA 91125, USA}

\altaffiltext{11}{Centro de Radioastronom\'{\i}a y Astrof\'{\i}sica, UNAM, 
	    Apartado Postal 3-72 (Xangari) 58089 Morelia, Michoacan, Mexico}

\altaffiltext{12}{Astronomy Department, University of Maryland, College Park, 
	    MD 20742, USA}


\begin{abstract}
We announce the discovery of SST-Lup3-1, a very low mass star close to
the brown dwarf boundary in Lupus III with a circum(sub)stellar disk,
discovered by the `Cores to Disks' Spitzer Legacy Program from mid-,
near-infrared and optical data, with very conspicuous crystalline
silicate features in its spectrum.  The Spectral Energy Distribution
of the system is fit with stellar and disk models to study the disk
structure.  SST-Lup3-1 is a M5.5 star with a luminosity of 0.08
L$_\odot$ (at an assumed distance of 200 pc), a mass of 0.10
M$_\odot$, a radius of 1.09 R$_\odot$, and an age of $\sim$ 1 Myrs.
It is the first of such objects with a full 5 to 35 $\mu$m spectrum
taken with the IRS and it shows strong 10 and 20\,$\mu$m silicate
features with high feature to continuum ratios and clear crystalline
features out to 33 $\mu$m. The mass of its flared disk is constrained
to lie between 2$\times10^{-4}$ and $10^{-7}$ M$_{\odot}$ (80--0.03
M$_\oplus$ masses) and the dust in the upper layer has a crystalline
silicate grain fraction between 15\% and 33\%, depending on the
assumed dust continuum. The availability of the full Spitzer infrared
spectrum allows an analysis of the dust composition as a function of
temperature and position in the disk. The hot ($\sim$300 K) dust
responsible for the 10\,$\mu$m feature consists of a roughly equal mix
of small ($\sim 0.1$ $\mu$m) and large ($\sim 1.5$ $\mu$m) grains,
whereas the cold ($\sim$70 K) dust responsible for the longer
wavelength silicate features contains primarily large grains ($\ge
1\,\mu$m). Since the cold dust emission arises from deeper layers in
the inner ($<$ 3 AU) disk as well as from the surface layers of the
outer (3-5 AU) disk, this provides direct evidence for combined grain
growth and settling in the disk. The inferred crystalline mass
fractions in the two components are comparable. Since only the inner
0.02 AU of the disk is warm enough to anneal the amorphous silicate
grains, even the lowest fraction of 15\% of crystalline material
requires either very efficient mixing or other formation mechanisms.
\end{abstract}


\keywords{ circumstellar disks --
                very low mass stars --
		stars: individual (SST-Lup3-1)
               }

\section{Introduction}
\label{intro}

Ground based observations of young brown dwarfs in star forming clouds
have shown the presence of significant infrared (IR) excesses
\citep[e.g.,][]{Comeron2000,Jayawardhana2003a} and broad H$\alpha$
emission lines \citep[e.g.,][]{Muzerolle2003,Natta2004} suggestive of
the presence of circum(sub)stellar disks and accretion in these very
low-mass objects. Also, recent ground-based mid-IR photometry has shown
hints of grain growth and dust settling from the silicate emission
features of some of these disks
\citep{Sterzik2004,Apai2004}. Observations with the Spitzer Space
Telescope (SST or Spitzer) permit the first detailed studies of the
geometry of such disks \citep[e.g.,][]{Allers2006} and of their
silicates \citep[e.g., ][]{Furlan2005} which provide interesting
comparisons with the dust evolution in disks around higher-mass
T Tauri and Herbig Ae/Be stars.

It is known that the presence and intensity of amorphous silicate
features in T Tauri stars anti-correlates with the size of the grains
in the disk surface layers (\citealt{Bouwman2001,vanBoekel2003},
\citeyear{vanBoekel2005}; \citealt{Przygodda2003};
\citealt{Kessler-Silacci2006}). \citet{Apai2005} analyzed Spitzer
spectra of the 10\,$\mu$m silicate feature of 6 brown dwarfs and very
low mass stars with disks in Chamaeleon I and found an anticorrelation
between the crystallinity of the silicates and the degree of disk
flaring, which suggests an evolutionary process in which the disk
flattens and the dust crystallizes as the grains grow and settle to
the disk mid-plane. This explanation, although very attractive, is
based on observations of only the 10\,$\mu$m silicate feature, with
low peak to continuum ratios and with spectra dominated by either
crystalline or amorphous features but not in combination.

In the present paper we announce the discovery of SST-Lup3-1, a new
very low mass star close to the brown dwarf boundary with a disk in
the Lupus III dark cloud, and analyze the properties of its disk from
its Spectral Energy Distribution (SED). Its Spitzer InfraRed
Spectrograph (IRS) spectrum over the entire 5 -- 35 $\mu$m range shows
strong amorphous and crystalline silicate emission features and is
presented and analyzed here to extract the mineralogical
composition of the dust in the disk.

\section{Target selection and observations}
\label{observations}

The `Cores to Disks' (c2d) Spitzer Legacy program \citep{Evans2003}
has mapped five nearby large molecular clouds with IRAC and MIPS at 6
wavelengths between 3.6 and 70 $\mu$m. Photometry was extracted for
all point sources and band-merged with the 2MASS catalog.  For more
information see the Delivery Documentation of the c2d project
\citep{Evans2006}. The area has also been surveyed with the 2.2m
telescope of the European Southern Observatory (ESO) and the Wide
Field Imager (WFI) instrument in the $R$, $I$, and $z$ broad-band
filters to complement the Spitzer data and study the very low-mass
stellar population (Comer\'on et al. in prep.). Detailed descriptions
of the observations of Lupus by the c2d Team can be found in Mundy et
al. (in prep.) for IRAC and in \citet{Chapman2006} for MIPS.

Using the band-merged catalog of the Lupus III dark cloud, we have
discovered a rich population of IR excess sources with luminosities in
the substellar domain according to the \citet{Baraffe1998} pre-main
sequence tracks. Constructing their SEDs with WFI $RIz$, 2MASS plus
IRAC and MIPS fluxes from 0.6 to 70 $\mu$m and comparing them with
stellar emission models, we find that some of the IR excesses at near
and mid IR wavelengths may be explained with the presence of
circum(sub)stellar disks. We selected one of them with optical and
near IR colors characteristic of a late M-type dwarf which could be
observed with the Spitzer IRS. The object, with coordinates 16$^{\rm
h}$11$^{\rm m}$59.9$^{\rm s}$, -38$^{\rm o}$23$'$37.5$''$ (J2000), is
named SSTc2d J161159.9-382337. We name it \object{SST-Lup3-1}
following \citet{Comeron2003}, and nickname it \object{Veronica's
star}.

SST-Lup3-1 was observed with the IRS onboard Spitzer on 14-08-2005,
with AOR Key 0015737856, in both short low (SL) and long low (LL)
modules. The total wavelength coverage is 5 -- 35 $\mu$m and the
integration times were 120 and 620 s for SL and LL, respectively. Data
reduction started from the Basic Calibrated Data images, pipeline
version S12.4.0. The processing includes bad-pixel correction,
extraction, defringing, and order matching using the c2d analysis
pipeline (\citealt{Kessler-Silacci2006,Lahuis2006}).

In addition, an optical spectrum of SST-Lup3-1 was obtained with the
EMMI instrument on ESO's New Technology Telescope. The object was
observed with the grism \# 6, which provides a spectral resolving
power of 1,500 with the 1$''$ slit, and with 9 minutes of integration
time. The spectrum was reduced using standard IRAF tasks for
extraction, wavelength calibration, and relative flux calibration
within the `onedspec' package.  Relative flux calibration was
performed by using the spectrum of the spectrophotometric standard
$\theta$ Vir as a reference.


\section{Stellar parameters}
\label{parameters}

Figure 1 shows the full optical spectrum of SST-Lup3-1 compared
to spectra of young M-type stars. It is clear that the object belongs
to an intermediate class between the M5 and M6 spectral types and that
the moderate extinction towards the source makes it resemble a colder
M6.5 brown dwarf. The TiO band-head absorptions at 6200, 7100 and 7600
\AA{} \citep{Kirkpatrick1991} and the slope of the continuum around
7500 \AA{} were used for the spectral typing after the extinction
correction.

We subsequently used an improved version of the method by
\citet{Frasca2003}, in which the spectrum is compared with a grid of
standard spectra of M to L dwarfs from \citet{Kirkpatrick1991} with
different extinctions to get the best fit to both parameters
simultaneously. This yields a spectral type of M5.5 with an error of
less than 0.5 spectral subclasses and an extinction of $A_V$=2.2
mag. A similar analysis with the method by \citet{Guieu2006}, which
uses instead an average of dwarf and giant M to L field standards to
better approach the emission of young stars, yields a spectral type of
M5.75 and $A_V$=2.08 mag. The similarity of these two results gives
confidence in our determination of the spectral type of the object.

The spectrum of SST-Lup3-1 shows H$\alpha$ (6563 \AA{}) in emission
with an equivalent width of 16.2 \AA{}. The profile is too narrow
(FWHM $\sim$ 5 \AA{}) to be a signature of disk accretion and is most
likely related to chromospheric activity (\citealt*{White2003},
\citealt{Jayawardhana2002}). However, its presence in the spectrum, as
well as in the other M6 and M6.5 young stars shown in the figure,
indicates the youth of the object and confirms its membership to the
Lupus molecular cloud. The detection of the Li {\sc i} absorption
line at 6708 \AA{} with an equivalent width of 0.5 \AA{} supports
the young nature of the object.

   \begin{figure}
     \label{EMMIspectrum}
   \centering
   \includegraphics[width=9cm]{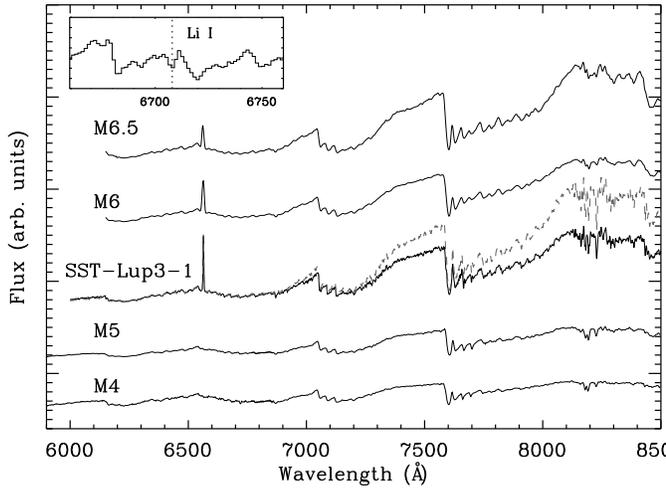}

   \caption{NTT/EMMI optical spectrum of SST-Lup3-1 compared to
   spectra of M4 (Gl 402) and M5 (GJ 1028) spectroscopic standards
   from \citet{Kirkpatrick1991} and M6 (Sz 112) and M6.5 (Sz 109)
   young dwarfs in Lupus III from \citet{Comeron2003}. The thin
   line corresponds to a dereddened spectrum with $A_V$=2.2 and the
   dashed line is the observed spectrum. The upper inset shows the Li
   {\sc i} absorption line at 6708 \AA{}.}

    \end{figure}

The SED of the object is constructed using the photometry in Table 1
and compared with StarDusty stellar models
\citep{Allard2000}. Following \citet{Luhman2003}, a young M5.5 star
has an effective temperature of 3057 K and intrinsic color $R_c-I_c$ =
1.88. Given the observed $R_c-I_c$ color of 2.00 we derive an
extinction from the optical photometry of $A_v = 0.67$ using the
extinction law of \citet*{Rieke1985}. The difference in the extinction
determinations from the optical photometry and spectrum is larger than
their uncertainties and cannot be due to the object having surface
gravity of a giant or dwarf. However it would disspaear if the star
would have an M5 spectral type, in which case $A_v \sim$ 2,
consistent with the spectroscopic result. Possible reasons for this
discrepancy are the scatter in the intrinsic $R_c-I_c$ colors of
M-type stars, that the object is variable at optical wavelengths, as
seen in other very low mass stars
\citep{Caballero2004}, that it presents bluer optical colors than
those expected for its spectral type as in \citet{Luhman1998}, or
that the standard interstellar extinction law is not valid for the
Lupus dark cloud.


   \begin{figure}
     \label{lupusBD2_SED}
   \centering
   \includegraphics[width=9cm]{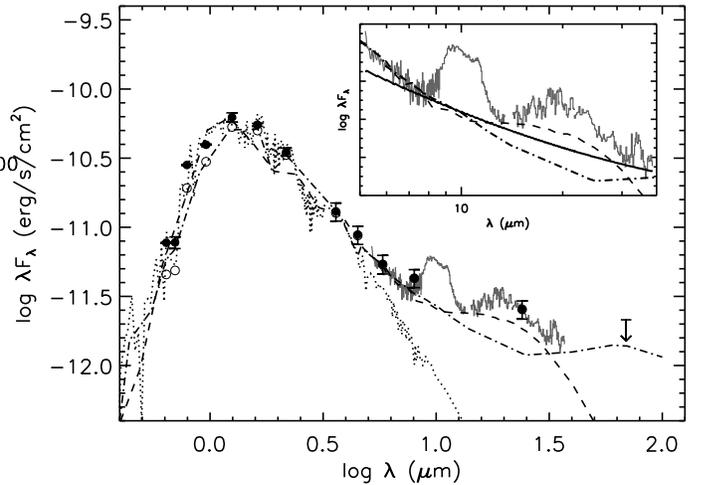}

   \caption{Spectral Energy Distribution of SST-Lup3-1, discovered in
   Lupus III with the c2d Spitzer data. The photometry is described in
   Table 1. Open and filled circles are the observed and dereddened
   fluxes respectively. The dotted line is the stellar model of a M5.5
   dwarf. The full Spitzer/IRS spectrum of the object is overplotted
   on the photometry and the total flux is fitted with a low mass CGPLUS
   flared disk model called 'high continuum' (dashed line) and a
   heavier D'Alessio et al. flared disk model called 'low continuum'
   (dot-dashed line) (see text). The inset illustrates the three
   different continua used to analyze the IRS spectrum; the full line
   is the usual power-law continuum.}

    \end{figure}
%


The difference in extinctions introduces only small differences in
luminosity of 0.015 L$_\odot$ and age of 0.4 Myrs so it is not
critical for the subsequent analysis. We thus adopt an extinction
value of $A_v$ = 0.7, which is consistent with all of the optical and
near-infrared photometry, and gives a luminosity of 0.081 L$_\odot$,
or log L/L$_\odot = -1.09$, for an assumed distance to Lupus III of
200 pc \citep{Comeron2006b}. Together with the $T_{\rm eff}$ of 3050
K, this gives an age of $\sim$ 1 Myr, a mass of 0.10 M$_\odot$ and a
radius of 1.09 R$_\odot$ using the \citet{Baraffe1998} isochrones.

\begin{table}
\label{photometry}
\caption{Photometry of SST-Lup3-1}             
\begin{tabular}{l l l l l }        
\hline\hline                 
Band      &  $\lambda$  & Magnitude & Flux  & Ref. \\
          &  $\mu$m     & mag & mJy    &  \\
\hline 
R       & 0.70 & 16.00 $\pm$ 0.10 &- & 1\\
R$_c$   & 0.64 & 16.251 $\pm$ 0.009 &- & 2\\
I$_c$   & 0.79 & 14.256 $\pm$ 0.007 &- & 2\\
z       & 0.96 & 13.597 $\pm$ 0.008 &- & 2\\
J$_{\rm 2MASS}$ & 1.25 &  12.197 $\pm$ 0.024  &- & 3\\
H$_{\rm 2MASS}$ & 1.62 &  11.511 $\pm$ 0.026  &- & 3\\
K$_{\rm 2MASS}$ & 2.20 &  11.204 $\pm$ 0.023  &- & 3\\
IRAC-1 &  3.60  &- &  17.03 $\pm$ 2.55   & 4\\
IRAC-2 & 4.50   &- &  14.60 $\pm$ 2.19   & 4\\
IRAC-3 & 5.80   &- &  11.70  $\pm$ 1.76   & 4\\
IRAC-4 &   8.00 &- &  12.90  $\pm$ 1.94   & 4\\
MIPS-1 &  24.0 &- &  22.40 $\pm$ 3.36   & 5\\
MIPS-2 &  70.0 &- &  $<$ 50.00     & 6\\
\hline                                   
\end{tabular}

References: 1. DENIS catalog (2005), 2. Comer\'on et al. (in prep.),
3. Cutri et al. (2003), 4. Mundy et al. (in prep.), 5. Chapman et
al. (2006), 6. Young et al. (2005)\\
\end{table}

\section{Disk parameters}
\label{diskparameters}

The aim of this work is to link the analysis of the SED, which is
related to the geometry of the disk, and the mid-infrared spectrum,
which gives the mineralogical composition of the dust in the upper
disk layer. The problem has no unique solution because the dust
composition is commonly determined by subtracting the disk continuum
SED from the IRS spectrum whereas the disk geometry and the SED itself
are determined partially by the dust properties. A larger sample of
$\sim$ 100 spectra of disk sources will be analyzed with the same
method (Olofsson, Augereau et al. in prep.); here we introduce the method
and the relevant ranges of disk and dust parameters for this specific
object.


We have computed flared irradiated disk models with the CGPLUS code
\citep{Dullemond2001} and flared irradiated accretion disk models with
the prescriptions of D'Alessio et al. (\citeyear{Dalessio2001},
\citeyear{Dalessio2005}) using the stellar parameters of SST-Lup3-1
as inputs for the central star. The dust compositions were chosen to
produce negligible silicate emission: pure carbon grains of
interstellar size in CGPLUS and large grains in the disk surface in
the D'Alessio et al.  models.  This provides possible disk continua to
subtract from the IRS spectrum assuming that the silicate features in
the spectrum are produced in an optically thin disk upper layer. A
model is considered a good fit for this analysis when it fits the
dereddened 2MASS and IRAC photometry, matches the IRS spectrum
shortward of 8 $\mu$m and does not exceed it at longer
wavelengths. This yield a range of disk parameters that fulfill that
condition: the CGPLUS disk models give an upper limit to the disk
mass (gas plus dust with the usual ratio of 100:1) of $10^{-7}$
M$_\odot$, inclination angle $i \ge 70 ^{\rm o}$ ($90 ^{\rm o}$ being
edge-on) and $R_{\rm disk} \le$ 15 AU. The D'Alessio et al. models
provide reasonable fits for disk masses between 2$\times10^{-4}$ and
0.018 M$_\odot$, inclinations $i \ge 30 ^{\rm o}$ and $R_{\rm disk}
\le$ 50 AU. The large difference in disk properties in the computed
cases comes mostly from the different dust opacities in the two sets
of models: the D'Alessio et al. models allow a range of maximum grain
sizes from 1 $\mu$m to 10 cm, which make the disk opacities low enough
to get very small infrared fluxes with larger disks than the CGPLUS
models, which use the dust opacity of astronomical carbon and silicate
sub-micron sized grains
\citep{Dullemond2001}. In any case, the quoted disk radii and masses
of these disk models are representative only of the material probed by
the IRS spectrum and therefore lower limits to the actual values. Both
sets of models produce flared disks with $H(R)/R
\propto R^{~0.2}$ and $\Sigma(R) \propto R^{-1}$.

Figure 2 shows the SED of SST-Lup3-1 together with example models,
with properties conveniently chosen to explore the range of reasonable
continua for the subsequent analysis of the IRS spectrum. The
dot-dashed line corresponds the `low continuum' model, a flared
D'Alessio et al. disk model with an outer disk radius of 10 AU, an
inner radius of 0.01 AU (3 times the stellar radius), a mass of
2.4$\times 10^{-4}$ M$_\odot$, a mass accretion rate of 10$^{-9}$
M$_\odot$yr$^{-1}$, a viscosity $\alpha$ = 0.01, a maximum grain size
of 100 $\mu$m and an inclination of 60$^{\rm o}$. It represents a
relatively massive disk with low dust opacities in this
comparison. The dashed line is the `high continuum' model, an
extremely low mass CGPLUS disk model with an outer radius of 15 AU, an
inner radius of 0.06 AU (20 times the stellar radius), a mass of
$10^{-7}$ M$_\odot$ and an inclination angle of 75$^{\rm o}$. More
massive disks with these small dust grains rapidly produce larger
fluxes than those seen in the IRS spectrum. The inset shows a blow-up
of the IRS spectrum with both continua together with the usual
power-law fit to the continuum \citep[as in][]{Kessler-Silacci2006}
for comparison. Both disk models (very low mass with very small dust
particles or more massive but with large dust particles) produce
similar SEDs up to the mid-IR wavelengths but have very different
shapes at long wavelengths. The need for large grains or very small
masses of small grains suggests that some degree of dust grain growth
and settling to the midplane have taken place in the disk of
SST-Lup3-1.

\section{Dust composition in the disk}
\label{irsanalysis}

\begin{figure}
   \centering
   \includegraphics[width=9cm]{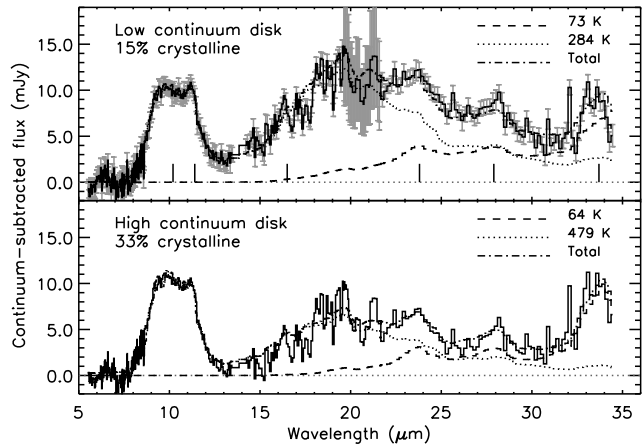}

   \caption{Continuum-subtracted Spitzer IRS spectrum of SST-Lup3-1
   and compositional analysis of the silicate emission. The thin line
   is the IRS spectrum of the source from which the low and high
   continuum disk models have been subtracted in the upper and lower
   panels respectively. The tickmarks identify the forsterite
   emission features. The upper panel shows the 1-$\sigma$ error
   bars. The best-fit models are shown in dot-dashed lines and
   combine the emission of amorphous and crystalline silicate grains
   with sizes of 0.1 and 1.5 $\mu$m of two different temperatures
   quoted in the plot. The fits show larger crystalline fractions for
   the case with high continuum and imply a minimum of 15\% of
   crystalline grains in the disk.}

   \label{compos_fit}%
    \end{figure}

Figure \ref{compos_fit} shows the IRS spectrum of SST-Lup3-1 from
which the disk models in Figure 2 have been subtracted. Both spectra
are then compared with the best-fit silicate emission model. The
spectrum has, to our knowledge, one of the highest peak to continuum
ratios ($\ge 2.0$) in the 10\,$\mu$m silicate feature ever published
for a disk around a very low mass star. The spectrum also shows
evidence of a moderately high degree of crystallization, with a
$S_{\rm 11.3}/S_{\rm 9.8}$ ratio equal to 0.9, and conspicuous
features from crystalline silicates such as forsterite at 11.4, 23.8,
27.9, and 33.7 $\mu$m \citep{Koike1993}. The 1-$\sigma$ error bars in
the upper panel quantify the signal to noise around these features.
Problems with the extraction of the third order of the LL module at
wavelengths between 19.5 and 21.5 $\mu$m for very faint sources do not
allow the identification of the 19.8 or 21.0 $\mu$m forsterite
features. The 10.2 and 16.5 $\mu $m features are not significant
within the noise. The rest of the crystalline features are present in
both full aperture and profile fit extractions, see \citet{Lahuis2006}
for more information about the quality assessment of the c2d IRS data
reduction pipeline.



%
%

\subsection{Method and results}

Compositional fits to the full IRS spectrum of SST-Lup3-1 have been
obtained following the method outlined in \citet{vanBoekel2005} and
used in \citet{Apai2005} and in \citet{Honda2006}: a $\chi^2$
minimization between the spectrum and a weighted sum of dust
opacities, multiplied by one or more blackbodies, for the five
principal dust species and for 2 grain sizes (0.1 $\mu$m and 1.5
$\mu$m). For every fixed blackbody temperature, this
procedure provides relative mass abundances, and a minimum $\chi^2$
value. This results in a relation between the temperatures and the
minimum $\chi^2$ values, from which we infer the best-fit
temperature. We use the distribution of hollow spheres routine for
calculating the opacities of the crystalline grains \citep{Min2005}
and the standard Mie theory for the amorphous grains.

Two novelties are applied here: the first one is that we use disk
models to estimate the disk continuum emission, which connects the
compositional analysis with the physical picture of the disk and
allows us to study the height and/or depth in the disk of the dust
emitting the 10 and 20\,$\mu$m amorphous and longer wavelength
crystalline silicate features. The second one is that we fit the
entire IRS spectrum from 5 -- 35 $\mu$m, which requires the inclusion
of an extra low-temperature dust component for the long wavelength
part and thus allows different parts of the disk to be probed.
This long wavelength range is also essential to ascertain the presence
of crystalline silicate features, many of which are potentially
confused with PAH features shortward of 13 $\mu$m. 

We first fit the 10\,$\mu$m silicate feature considering only the
spectral range between 7.5 and 13.5 $\mu$m, which is least affected by
the continuum choice and which requires a relatively high temperature
of $\sim$300 K. This best fit is then subtracted from the spectrum and
the residual spectrum is fit with a lower temperature component with a
best-fit value of $\sim$70 K.  This method has the virtue that it
allows a direct comparison with fits performed to only the 10 $\mu$m
feature (as in Apai et al. 2005, van Boekel et al. 2005), where only
the high temperature component is considered. It also allows a direct
determination of the mass fractions in each component, which are 59
and 41\% for the hot and cold components respectively if the low
continuum model is used. With the high continuum, these numbers change
to 65 and 35\%. Also, the two temperature approach identifies (via
the $\chi^2$ minimization) the characteristic temperature ranges of
the dust producing the amorphous 10 and 20\,$\mu$m silicate features.


\begin{table*}
\label{bd2composition}
\caption{Mineralogical composition of the disk around SST-Lup3-1 after
subtraction of the low continuum disk model.}

\begin{tabular}{l | l l | l l}           
\hline\hline                 
 Species     &  Mass \%$^a$  &  & Mass \%$^a$ &  \\
             & 284$\pm$27 K & &  73$\pm$9 K &  \\
             &  0.1 $\mu$m & 1.5 $\mu$m &  0.1 $\mu$m & 1.5 $\mu$m \\
\hline 
Amorphous silicates$^b$ & 33.5$\pm$2.7 & 49.8$\pm$5.3 & 0.0+1.5 & 87.8$\pm$9.6 \\ 
Crystalline forsterite & 6.0$\pm$0.5 & 2.3$\pm$0.2 & 11.8$\pm$1.2  & 0.2$\pm$0.1 \\
Crystalline enstatite  & 0.7$\pm$0.1 & 7.7$\pm$0.6 & 0.0+1.5 & 0.2$\pm$0.1 \\
\hline                                   
\end{tabular}

$^a$ Relative fractions of 59\% (284 K) and 41\% (73 K).\\
$^b$ Includes olivine, pyroxene and silica.
\end{table*}

Table 2 shows the abundances of the different dust species that result
in the best fit to the IRS spectrum shown in Figure 3 after
subtracting the low continuum disk model. The abundances are
found by averaging a number of realizations of the fit in which the
spectral ranges used for the fits were changed and in which a range of
temperatures was explored. The associated uncertainties are therefore
larger than indicated by the standard deviations of the data, since
they account for the relatively low number of realizations. However,
they are similar to those cited in \cite{Apai2005} and in
\cite{Honda2006} in their fits to the 10\,$\mu$m spectral range (high
temperature component). The uncertainties in derived abundances are
slightly larger for the longer wavelength spectral range.  The
quantitative fit results obviously depend sensitively on the adopted
continuum, but the qualitative results are very similar. Hence, we
will discuss primarily the low continuum case below, and mention the
high continuum case only when it gives a different trend.  Also, Table
2 only provides a combined mass percentage for the amorphous
silicates, which includes olivine, pyroxene and silica, because the
fitting method was not sufficiently robust to identify the individual
amorphous species responsible for the long-wavelength part of the
spectrum. This does not affect the relative mass fraction of
amorphous and crystalline material, which was always consistently
recovered.

The best fit composition shows that i) the amorphous hot silicates are
a roughly equal mix of small and large grains, which is also found in
other sources with high peak to continuum ratios ($\sim$2.0) of the
10\,$\mu$m silicate feature (see e.g. Fig. 3 in \citealt{Honda2006});
ii) the crystalline silicates in the hot component are a mix of large
and small grains independently of the chosen continuum, a result also
consistent with that given by \citet{vanBoekel2005} for similar
10\,$\mu$m features; iii) the amorphous cold grains are mostly large
(1.5 $\mu$m). This is a new and surprising result. However, this
conclusion depends on the adopted continuum, since the high continuum
yields mostly small grains in the cold component; iv) the cold
crystalline grains are found to be smaller than 2\,$\mu$m regardless
of the continuum used.

In the following, we discuss in detail the total crystalline mass
fraction and the distribution of large and small grains in the disk.

\subsection{Crystalline mass fraction}

The best-fit crystalline mass percentage varies from 33\% for the high
disk continuum, through 20\% with the power law continuum and down to
15\% with the low disk continuum. In general, it is found that the
smaller the peak to continuum ratio, the larger the mass crystalline
fraction. The case with the low continuum based on the D'Alessio et
al.\ (2005) models represents the best fit to the spectrum, mostly at
10\,$\mu$m, and also the lower limit to the total dust crystalline
fraction, therefore all subsequent analyses will focus on this case.

The percentages of crystalline grains in the hot and cold components
are 16.7 and 12.2 \% respectively, with a similar ratio but larger
absolute values in the high continuum case. The clear identification
of the 23.7, 27.9 and 33.7 $\mu$m forsterite features in the long
wavelength spectrum is a good assessment of the presence of
crystalline grains in the disk and confirms the result at short
wavelengths. 
We have also compared our results to those of \citet{Apai2005} for
sources with similar 10\,$\mu$m feature shapes using the high
continuum case, which is the closest approximation to the definition
of the continuum of those authors.  Our inferred crystalline mass
fractions are similar to better than 10\%.

\subsection{Distribution of large and small grains}

Regardless of the adopted continuum, our fitting technique of the
hot component recovers the approximately 50--50\% of small (0.1
$\mu$m) and large (1.5 $\mu$m) grains found by previous analyses for
this type of 10\,$\mu$m feature. The fit to the long wavelength part
of the spectrum yields mostly large grains for the low continuum (87.7
\% of large grains, Table \ref{bd2composition}) and mostly small
grains for the high continuum (76.3 \% of small grains).

To test the sensitivity of the results to the adopted large grain size, a
series of fits to the long wavelength range of the IRS spectrum
was performed with 1.0 and 5.0 $\mu$m grain sizes but this did
not improve the fits dramatically. The results were consistent with
what has been reported above, namely that most of the grains in the
low-temperature component are 1 $\mu$m-sized amorphous grains for the
low continuum case. Therefore, even with the small difference in
feature profiles for grains in the 1--5 $\mu$m size range, the best
fit always selects grains of $\sim$1\,$\mu$m radius for the low
temperature component.

The crystalline grains are also a mix of large and small grains in the
hot component regardless of the continuum used. However, in the tests
with larger grain sizes for the cold component, we find that the best
fit always selects the smallest available grains, with grains of 0.1
-- 1.5\,$\mu$m giving the best fit to the long wavelength spectrum. In
these cases, the cold temperature components of the silicate emission
model (dashed lines in Fig. 3) show the $\sim$1:1:2 relative flux
ratios between the 23.7, 27.9 and 33.7 $\mu$m forsterite features for
small grains (Fig. 8 in \citealt{Kessler-Silacci2006}).


\section{Discussion}

\begin{figure}[h]
   \centering
   \includegraphics[width=9cm]{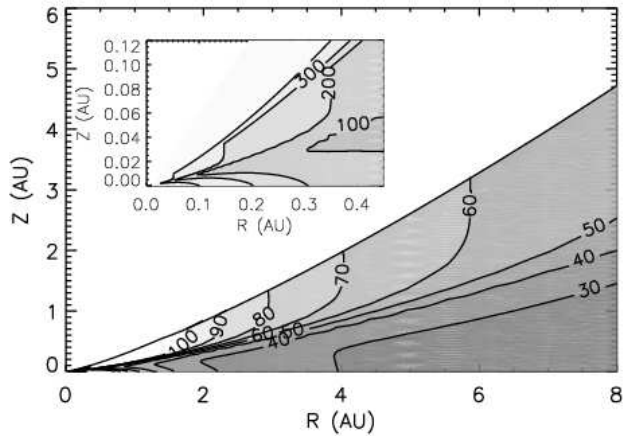}

   \caption{Two dimensional distribution of temperature of the
   irradiated accretion disk model used to fit the SED of SST-Lup3-1,
   based on the D'Alessio et al. models, illustrating the different
   emission zones. The gas and dust temperatures of the disk are given
   in Kelvin and shown as contours. The inset shows in detail the
   hotter upper layers in the inner disk, not visible in the larger
   figure due to their small physical scale. The cold ($\sim$70 K)
   emission component extends from 5 to 0.7 AU and reaches the half
   disk height in the innermost edge. The hot ($\sim$300 K) component
   is mostly localized in the inner disk and surface layers up to 1
   AU.}

   \label{disk_struc}%
    \end{figure}

In this section we connect the results of the compositional
analysis with the disk modeling to provide a combined interpretation
of all the data. As shown in Table 2, the hot temperature emission
component contributes 59\% of the total dust mass, reproduces
completely the 10\,$\mu$m feature and contributes to the 20\,$\mu$m
feature. The cold temperature component, mostly coming from large
amorphous grains, reproduces the rest of the 20\,$\mu$m feature and
accounts for most of the forsterite features longward than
20\,$\mu$m. Figure
\ref{disk_struc} shows the 2D temperature structure of the 
D'Alessio et al.\ disk model (from Sect. 4) used to
fit the entire SED except the silicate emission features. 
The $\sim$300 and $\sim$70~K contours indicate
the locations in the disk surface where the different features
are mostly produced. The hot inner disk region (from
0.05 to 1 AU) with temperatures around 300 K emits mostly in the
10 $\mu$m range. The longer wavelength emission comes from dust at
a temperature of around 70 K, which is found both in the outer disk
surface (from $\sim$3 to 5 AU) and in the deeper disk layers at smaller
radii.

Interestingly, the majority of the grains in the low temperature
component are large grains, while only half of the grains in the high
temperature component are large. Together with the distribution of
dust at 70 K in the disk, this strengthens the possibility that part
of the long wavelength emission spectrum comes from {\sl deeper
layers} of the inner disk rather than just the outer disk
surface. This could be interpreted as a signature of sedimentation to
the disk mid-plane.
This conclusion is not affected by optical depth: the total gas plus
dust density in most of those layers (from the disk surface to 1/2 of
the disk's height) is not high enough to render the disk optically
thick.
This would naturally explain the presence of larger grains in the
low-temperature component, and is consistent with the disk model used
for the underlying continuum which assumes dust grains as large
as 100\,$\mu$m in size in the disk interior (Sect.\  4).


Concerning the crystalline silicates, their mass percentages in
the hot and cold components are 16.7\% and 12.2\% respectively. Given
that 59\% of the dust mass is in the hot component, this means that
there are $\sim$2 times more crystalline grains in the inner disk than
in the outer disk. This is consistent with the radial distribution of
crystalline material reported by the high spatial resolution
observations in the inner disks in Herbig Ae/Be stars
\citep{vanBoekel2004} and suggests similar phenomena in disks around
the very low mass stars.

Van Boekel et~al. \citeyear{vanBoekel2005} and \citet{Apai2005} report
correlations between high crystalline fractions and overall large
grain sizes or flatter disks respectively, based on 10\,$\mu$m
spectra only. Both results suggest a relation between the
crystallization and grain growth processes. Our results for the
hot component alone are consistent with those trends, however, the
complete 5 to 35 $\mu$m spectrum of SST-Lup3-1 shows a much more
complex scenario: higher amounts of crystalline silicates in the hot
component, but not necessarily linked with larger amorphous grains,
and mostly large amorphous grains in the cold component, where the
crystalline fraction is only slightly smaller.



Here we speculate on a scenario implied by the observed dust
distributions: the inner disk is a turbulent region where dust is
being partially annealed when the temperature is above 800 K and where
grain growth is efficient for all particles because of the higher
densities \citep{Kessler-Silacci2007}. The outer disk contains small
particles with a non-negligible percentage of crystalline material
which has either been transported out radially \citep{Gail2004} or
which have formed via a different mechanism than annealing. The
10\,$\mu$m feature comes mostly from the hot and dense inner disk down
to a depth in the disk where both large and small grains are
probed. The long wavelength part of the spectrum is emitted by large
amorphous grains in deep layers of the disk at distances from 0.7 to 5
AU, and by small crystalline grains in the disk surface further than 3
AU. According to this, the crystalline material could be
preferentially found in the disk surface and in the hot part of the
innermost disk and the larger amorphous grains would be found closer
to the midplane once grain growth and settling have taken place. Then,
the high crystalline fractions in disks with large grains could be
explained if the lower dust opacities (due to the larger grains) would
allow the 10\,$\mu$m feature to probe deeper in the inner disk, where
the large and crystalline grains are abundant, as in
\citet{vanBoekel2005}. Such disks would also consistently have a
smaller flaring angle, as found by \citet{Apai2005}. On the other
hand, disks where grain growth has not occurred may have small dust
with large opacities in the inner disks, large flaring angles and a
10\,$\mu$m feature coming from the surface of the inner disk, where
the abundance of large grains is smaller.

Finally, even assuming the most conservative estimate of the disk
crystalline mass fraction of 15\%, the disk of SST-Lup3-1 shows a
remarkably high degree of crystallinity taking into account that only
0.2\% of the disk (from 0.01 to 0.2 AU, out of 10 AU) is warm enough
to produce crystalline material. Three main phenomena may explain this
result: i) there is another source of disk heating such as X-ray
flares \citep[e.g., ][]{Preibisch2005} or shocks that are operating in the
system, ii) there was an episodic heating event e.g., due to a sudden
increase in the disk mass accretion rate as observed in FU Ori stars
\citep{Hartmann1996} or iii) there is efficient radial and vertical
transport of dust in the system. \citet{Gail2004} show that radial
mixing in disks can produce crystallinity fractions of several tens of
per cent at distances of 5 to 10 AU, although that model predicts
larger amounts of forsterite grains in the inner than outer disk,
contrary to our observations.  Future X-ray observations of the
source, dynamically self-consistent disk models of the object, and a
larger sample of complete 5 to 35 $\mu$m IRS spectra of disks may help
elucidate which is the main effect at work.

\section{Conclusions}
\label{conclussion}

A detailed study of the disk emission of SST-Lup3-1 is presented and
the results obtained are as follows:

\begin{itemize}

\item {SST-Lup3-1 is a newly discovered very low mass M5.5 young star
close to the hydrogen burning boundary with a disk, found with the
Spitzer c2d data in the Lupus III dark cloud.}

\item {The complete 5--35 $\mu$m mid-IR Spitzer/IRS spectrum of such a
low luminosity object is presented for the first time. It contains one
of the highest 10 and 20\,$\mu$m peak to continuum ratios reported for
such a source. }

\item {A compositional fit of the spectrum shows two temperature
components. A hot (T $\sim$ 300 K) component with a mix of small and
large grains and a cold component (T $\sim$ 70 K) with mostly large
grains. This latter conclusion depends on the adopted continuum.}

\item {The spectrum has prominent spectral signatures indicating
significant amounts of crystalline grains in the disk, confirmed in
the long wavelength range of the spectrum. The crystalline mass
fraction in the cold component is only slightly less than that in the
hot disk component.  }

\item {The crystalline grains in the hot component are a mix
of large (1.5 $\mu$m) and small (0.1 $\mu$m) grains, while those in
the cold component are mostly small grains, suggesting that the 
crystalline material is preferentially in the disk surface layers.}

\item{The cold component emission can arise from both the ``outer''
disk surface or from deeper layers in the inner disk. In the latter
case, this would be direct evidence for combined grain growth and
settling with vertical height.}

\end{itemize}

\acknowledgements
     
B. M. thanks the Fundaci\'on Ram\'on Areces for financial support. The
authors thank Michiel Min and to Jeroen Bouwman for fruitful
discussions on crystalline silicates and the referee for a detailed
report which improved the impact of the paper substantially. Support
for this work, part of the Spitzer Space Telescope Legacy Science
Program, was provided through Contract Numbers 1256316, 1224608 and
1230780 issued by the Jet Propulsion Laboratory, California Institute
of Technology under NASA contract 1407. Astrochemistry at Leiden is
supported by a NWO Spinoza and NOVA grant, and by the European
Research Training Network ``The Origin of Planetary Systems''
(PLANETS, contract number HPRN-CT-2002-00308).

\bibliographystyle{aastex}
\bibliography{lupus_bd2.bbl}

\begin{thebibliography}{42}
\expandafter\ifx\csname natexlab\endcsname\relax\def\natexlab#1{#1}\fi

\bibitem[{{Allard} {et~al.}(2000){Allard}, {Hauschildt}, \&
  {Schweitzer}}]{Allard2000}
{Allard}, F., {Hauschildt}, P.~H., \& {Schweitzer}, A. 2000, \apj, 539, 366

\bibitem[{{Allers} {et~al.}(2006){Allers}, {Kessler-Silacci}, {Cieza}, \&
  {Jaffe}}]{Allers2006}
{Allers}, K.~N., {Kessler-Silacci}, J.~E., {Cieza}, L.~A., \& {Jaffe}, D.~T.
  2006, \apj, 644, 364

\bibitem[{{Apai} {et~al.}(2005){Apai}, {Pascucci}, {Bouwman}, {Natta},
  {Henning}, \& {Dullemond}}]{Apai2005}
{Apai}, D., {Pascucci}, I., {Bouwman}, J., {et~al.} 2005, Science, 310, 834

\bibitem[{{Apai} {et~al.}(2004){Apai}, {Pascucci}, {Sterzik}, {van der Bliek},
  {Bouwman}, {Dullemond}, \& {Henning}}]{Apai2004}
{Apai}, D., {Pascucci}, I., {Sterzik}, M.~F., {et~al.} 2004, \aap, 426, L53

\bibitem[{{Baraffe} {et~al.}(1998){Baraffe}, {Chabrier}, {Allard}, \&
  {Hauschildt}}]{Baraffe1998}
{Baraffe}, I., {Chabrier}, G., {Allard}, F., \& {Hauschildt}, P.~H. 1998, \aap,
  337, 403

\bibitem[{{Bouwman} {et~al.}(2001){Bouwman}, {Meeus}, {de Koter}, {Hony},
  {Dominik}, \& {Waters}}]{Bouwman2001}
{Bouwman}, J., {Meeus}, G., {de Koter}, A., {et~al.} 2001, \aap, 375, 950

\bibitem[{{Caballero} {et~al.}(2004){Caballero}, {B{\'e}jar}, {Rebolo}, \&
  {Zapatero Osorio}}]{Caballero2004}
{Caballero}, J.~A., {B{\'e}jar}, V.~J.~S., {Rebolo}, R., \& {Zapatero Osorio},
  M.~R. 2004, \aap, 424, 857

\bibitem[{{Chapman} {et~al.}(2006){Chapman}, {Evans}, {Allen}, {Blake},
  {Boogert}, {Bourke}, {Harvey}, {Kessler}, {Koerner}, {Lee}, {Mundy}, {Myers},
  {Padgett}, {Pontoppidan}, {Sargent}, {Stapelfeldt}, {van Dishoeck}, {Young},
  \& {Young}}]{Chapman2006}
{Chapman}, N., {Evans}, II, N.~J., {Allen}, L.~E., {et~al.} 2006, \apj,
  submitted

\bibitem[{{Comer{\'o}n}(2006)}]{Comeron2006b}
{Comer{\'o}n}, F. 2006, {The Lupus Dark Clouds}, ed. B.~{Reipurth} (Handbook of
  Low Mass Star Formation in Southern Molecular Clouds), submitted

\bibitem[{{Comer{\'o}n} {et~al.}(2003){Comer{\'o}n}, {Fern{\'a}ndez},
  {Baraffe}, {Neuh{\"a}user}, \& {Kaas}}]{Comeron2003}
{Comer{\'o}n}, F., {Fern{\'a}ndez}, M., {Baraffe}, I., {Neuh{\"a}user}, R., \&
  {Kaas}, A.~A. 2003, \aap, 406, 1001

\bibitem[{{Comer{\'o}n} {et~al.}(2000){Comer{\'o}n}, {Neuh{\"a}user}, \&
  {Kaas}}]{Comeron2000}
{Comer{\'o}n}, F., {Neuh{\"a}user}, R., \& {Kaas}, A.~A. 2000, \aap, 359, 269

\bibitem[{{D'Alessio} {et~al.}(2001){D'Alessio}, {Calvet}, \&
  {Hartmann}}]{Dalessio2001}
{D'Alessio}, P., {Calvet}, N., \& {Hartmann}, L. 2001, \apj, 553, 321

\bibitem[{{D'Alessio} {et~al.}(2005){D'Alessio}, {Mer{\'{\i}}n}, {Calvet},
  {Hartmann}, \& {Montesinos}}]{Dalessio2005}
{D'Alessio}, P., {Mer{\'{\i}}n}, B., {Calvet}, N., {Hartmann}, L., \&
  {Montesinos}, B. 2005, Revista Mexicana de Astronomia y Astrofisica, 41, 61

\bibitem[{{Dullemond} {et~al.}(2001){Dullemond}, {Dominik}, \&
  {Natta}}]{Dullemond2001}
{Dullemond}, C.~P., {Dominik}, C., \& {Natta}, A. 2001, \apj, 560, 957

\bibitem[{{Evans} {et~al.}(2003){Evans}, {Allen}, {Blake}, {Boogert}, {Bourke},
  {Harvey}, {Kessler}, {Koerner}, {Lee}, {Mundy}, {Myers}, {Padgett},
  {Pontoppidan}, {Sargent}, {Stapelfeldt}, {van Dishoeck}, {Young}, \&
  {Young}}]{Evans2003}
{Evans}, II, N.~J., {Allen}, L.~E., {Blake}, G.~A., {et~al.} 2003, \pasp, 115,
  965

\bibitem[{{Evans} {et~al.}(2006){Evans}, {Harvey}, {Dunham}, {Mundy}, {Lai},
  {Chapman}, {Huard}, {Brooke}, \& {Koerner}}]{Evans2006}
{Evans}, II, N.~J., {Harvey}, P.~M., {Dunham}, M.~M., {et~al.} 2006,
  http://ssc.spitzer.caltech.edu/legacy/original.html

\bibitem[{{Frasca} {et~al.}(2003){Frasca}, {Alcal{\'a}}, {Covino}, {Catalano},
  {Marilli}, \& {Paladino}}]{Frasca2003}
{Frasca}, A., {Alcal{\'a}}, J.~M., {Covino}, E., {et~al.} 2003, \aap, 405, 149

\bibitem[{{Furlan} {et~al.}(2005){Furlan}, {Calvet}, {D'Alessio}, {Hartmann},
  {Forrest}, {Watson}, {Luhman}, {Uchida}, {Green}, {Sargent}, {Najita},
  {Sloan}, {Keller}, \& {Herter}}]{Furlan2005}
{Furlan}, E., {Calvet}, N., {D'Alessio}, P., {et~al.} 2005, \apjl, 621, L129

\bibitem[{{Gail}(2004)}]{Gail2004}
{Gail}, H.-P. 2004, \aap, 413, 571

\bibitem[{{Guieu} {et~al.}(2006){Guieu}, {Dougados}, {Monin}, {Magnier}, \&
  {Mart{\'{\i}}n}}]{Guieu2006}
{Guieu}, S., {Dougados}, C., {Monin}, J.-L., {Magnier}, E., \& {Mart{\'{\i}}n},
  E.~L. 2006, \aap, 446, 485

\bibitem[{{Hartmann} \& {Kenyon}(1996)}]{Hartmann1996}
{Hartmann}, L. \& {Kenyon}, S.~J. 1996, \araa, 34, 207

\bibitem[{{Honda} {et~al.}(2006){Honda}, {Kataza}, {Okamoto}, {Yamashita},
  {Min}, {Miyata}, {Sako}, {Fujiyoshi}, {Sakon}, \& {Onaka}}]{Honda2006}
{Honda}, M., {Kataza}, H., {Okamoto}, Y.~K., {et~al.} 2006, \apj, 646, 1024

\bibitem[{{Jayawardhana} {et~al.}(2002){Jayawardhana}, {Mohanty}, \&
  {Basri}}]{Jayawardhana2002}
{Jayawardhana}, R., {Mohanty}, S., \& {Basri}, G. 2002, \apjl, 578, L141

\bibitem[{{Jayawardhana} {et~al.}(2003){Jayawardhana}, {Mohanty}, \&
  {Basri}}]{Jayawardhana2003a}
{Jayawardhana}, R., {Mohanty}, S., \& {Basri}, G. 2003, \apj, 592, 282

\bibitem[{{Kessler-Silacci} {et~al.}(2006){Kessler-Silacci}, {Augereau},
  {Dullemond}, {Geers}, {Lahuis}, {Evans}, {van Dishoeck}, {Blake}, {Boogert},
  {Brown}, {J{\o}rgensen}, {Knez}, \& {Pontoppidan}}]{Kessler-Silacci2006}
{Kessler-Silacci}, J., {Augereau}, J.-C., {Dullemond}, C.~P., {et~al.} 2006,
  \apj, 639, 275

\bibitem[{{Kessler-Silacci} {et~al.}(2007){Kessler-Silacci}, {Dullemond},
  {Augereau}, {Geers}, {van Dishoeck}, {Evans}, {Blake}, \&
  {Brown}}]{Kessler-Silacci2007}
{Kessler-Silacci}, J., {Dullemond}, C.~P., {Augereau}, J.-C., {et~al.} 2007,
  \apj, submitted

\bibitem[{{Kirkpatrick} {et~al.}(1991){Kirkpatrick}, {Henry}, \&
  {McCarthy}}]{Kirkpatrick1991}
{Kirkpatrick}, J.~D., {Henry}, T.~J., \& {McCarthy}, Jr., D.~W. 1991, \apjs,
  77, 417

\bibitem[{{Koike} {et~al.}(1993){Koike}, {Shibai}, \& {Tuchiyama}}]{Koike1993}
{Koike}, C., {Shibai}, H., \& {Tuchiyama}, A. 1993, \mnras, 264, 654

\bibitem[{{Lahuis} {et~al.}(2006){Lahuis}, {Kessler-Silacci}, {Evans}, {Blake},
  {van Dishoeck}, {Augereau}, {Bandihi}, {Boogert}, {Brown}, {Geers},
  {Joergensen}, {Knez}, {Mer\'{\i}n}, {Olofsson}, \& M.}]{Lahuis2006}
{Lahuis}, F., {Kessler-Silacci}, J., {Evans}, N., {et~al.} 2006, {c2d
  Spectroscopy Explanatory Supplement}, Tech. rep., Pasadena: Spitzer Science
  Center

\bibitem[{{Luhman} {et~al.}(2003){Luhman}, {Brice{\~n}o}, {Stauffer},
  {Hartmann}, {Barrado y Navascu{\'e}s}, \& {Caldwell}}]{Luhman2003}
{Luhman}, K.~L., {Brice{\~n}o}, C., {Stauffer}, J.~R., {et~al.} 2003, \apj,
  590, 348

\bibitem[{{Luhman} {et~al.}(1998){Luhman}, {Briceno}, {Rieke}, \&
  {Hartmann}}]{Luhman1998}
{Luhman}, K.~L., {Briceno}, C., {Rieke}, G.~H., \& {Hartmann}, L. 1998, \apj,
  493, 909

\bibitem[{{Min} {et~al.}(2005){Min}, {Hovenier}, \& {de Koter}}]{Min2005}
{Min}, M., {Hovenier}, J.~W., \& {de Koter}, A. 2005, \aap, 432, 909

\bibitem[{{Muzerolle} {et~al.}(2003){Muzerolle}, {Hillenbrand}, {Calvet},
  {Brice{\~n}o}, \& {Hartmann}}]{Muzerolle2003}
{Muzerolle}, J., {Hillenbrand}, L., {Calvet}, N., {Brice{\~n}o}, C., \&
  {Hartmann}, L. 2003, \apj, 592, 266

\bibitem[{{Natta} {et~al.}(2004){Natta}, {Testi}, {Muzerolle}, {Randich},
  {Comer{\'o}n}, \& {Persi}}]{Natta2004}
{Natta}, A., {Testi}, L., {Muzerolle}, J., {et~al.} 2004, \aap, 424, 603

\bibitem[{{Preibisch} {et~al.}(2005){Preibisch}, {McCaughrean}, {Grosso},
  {Feigelson}, {Flaccomio}, {Getman}, {Hillenbrand}, {Meeus}, {Micela},
  {Sciortino}, \& {Stelzer}}]{Preibisch2005}
{Preibisch}, T., {McCaughrean}, M.~J., {Grosso}, N., {et~al.} 2005, \apjs, 160,
  582

\bibitem[{{Przygodda} {et~al.}(2003){Przygodda}, {van Boekel},
  {{\`A}brah{\`a}m}, {Melnikov}, {Waters}, \& {Leinert}}]{Przygodda2003}
{Przygodda}, F., {van Boekel}, R., {{\`A}brah{\`a}m}, P., {et~al.} 2003, \aap,
  412, L43

\bibitem[{{Rieke} \& {Lebofsky}(1985)}]{Rieke1985}
{Rieke}, G.~H. \& {Lebofsky}, M.~J. 1985, \apj, 288, 618

\bibitem[{{Sterzik} {et~al.}(2004){Sterzik}, {Pascucci}, {Apai}, {van der
  Bliek}, \& {Dullemond}}]{Sterzik2004}
{Sterzik}, M.~F., {Pascucci}, I., {Apai}, D., {van der Bliek}, N., \&
  {Dullemond}, C.~P. 2004, \aap, 427, 245

\bibitem[{{van Boekel} {et~al.}(2004){van Boekel}, {Min}, {Leinert}, {Waters},
  {Richichi}, {Chesneau}, {Dominik}, {Jaffe}, {Dutrey}, {Graser}, {Henning},
  {de Jong}, {K{\"o}hler}, {de Koter}, {Lopez}, {Malbet}, {Morel}, {Paresce},
  {Perrin}, {Preibisch}, {Przygodda}, {Sch{\"o}ller}, \&
  {Wittkowski}}]{vanBoekel2004}
{van Boekel}, R., {Min}, M., {Leinert}, C., {et~al.} 2004, \nat, 432, 479

\bibitem[{{van Boekel} {et~al.}(2005){van Boekel}, {Min}, {Waters}, {de Koter},
  {Dominik}, {van den Ancker}, \& {Bouwman}}]{vanBoekel2005}
{van Boekel}, R., {Min}, M., {Waters}, L.~B.~F.~M., {et~al.} 2005, \aap, 437,
  189

\bibitem[{{van Boekel} {et~al.}(2003){van Boekel}, {Waters}, {Dominik},
  {Bouwman}, {de Koter}, {Dullemond}, \& {Paresce}}]{vanBoekel2003}
{van Boekel}, R., {Waters}, L.~B.~F.~M., {Dominik}, C., {et~al.} 2003, \aap,
  400, L21

\bibitem[{{White} \& {Basri}(2003)}]{White2003}
{White}, R.~J. \& {Basri}, G. 2003, \apj, 582, 1109

\end{thebibliography}


\end{document}